\documentclass[aps,prb,reprint,superscriptaddress,floatfix,showpacs]{revtex4-1}
\usepackage{graphicx}
\usepackage{dcolumn} 
\usepackage{bm}      
\usepackage[utf8]{inputenc}
\usepackage{color}
\usepackage{hyperref}
\usepackage{siunitx}
\usepackage{amsmath}
\usepackage{ulem}
\usepackage{float}
\usepackage{xfrac}

\definecolor{purple}{RGB}{128,0,128}

\let\vec\mathbf

\date{\today}

\begin{document}

\title{Observation of Kekulé vortices induced in graphene by hydrogen adatoms}

\author{Yifei Guan}
\affiliation{Institute of Physics, École Polytechnique Fédérale de Lausanne (EPFL), CH-1015 Lausanne, Switzerland}

\author{C. Dutreix}
\affiliation{Univ. Bordeaux, CNRS, LOMA, UMR 5798, F-33400 Talence, France}
 
\author{H. Gonz\'{a}lez-Herrero}
\affiliation{Departamento de F\'isica de la Materia Condensada, Universidad Aut\'onoma de Madrid, E-28049 Madrid, Spain}
\affiliation{Condensed Matter Physics Center (IFIMAC), Universidad Aut\'onoma de Madrid, E-28049 Madrid, Spain}

\author{M. M. Ugeda}
\affiliation{Donostia International Physics Center (DIPC), Paseo Manuel de Lardizábal 4, 20018 San Sebastián, Spain}
\affiliation{Centro de Física de Materiales (CSIC-UPV-EHU), Paseo Manuel de Lardizábal 5, 20018 San Sebastián, Spain}
\affiliation{Ikerbasque, Basque Foundation for Science, 48013 Bilbao, Spain}
	
\author{I.~Brihuega}
\affiliation{Departamento de F\'isica de la Materia Condensada, Universidad Aut\'onoma de Madrid, E-28049 Madrid, Spain}
\affiliation{Condensed Matter Physics Center (IFIMAC), Universidad Aut\'onoma de Madrid, E-28049 Madrid, Spain}
\affiliation{Instituto Nicol\'as Cabrera, Universidad Aut\'onoma de Madrid, E-28049 Madrid, Spain}

\author{M. I. Katsnelson}
\affiliation{Institute for Molecules and Materials, Radboud University, Heijendaalseweg 135, 6525AJ Nijmegen, The Netherlands}

\author{O. V. Yazyev}
\email{email: oleg.yazyev@epfl.ch}
\affiliation{Institute of Physics, École Polytechnique Fédérale de Lausanne (EPFL), CH-1015 Lausanne, Switzerland}

\author{Vincent T. Renard}
\email{email: vincent.renard@cea.fr}
\affiliation{Univ. Grenoble Alpes, CEA, Grenoble INP, IRIG, PHELIQS, 38000 Grenoble, France}

\maketitle

Fractional charges are one of the wonders of the fractional quantum Hall effect, a liquid of strongly correlated electrons in a large magnetic field\cite{Laughlin1983,dePicciotto1997,Saminadayar1997}. Fractional excitations are also anticipated in two-dimensional crystals of non-interacting electrons under time-reversal symmetry, as bound states of a rotating bond order known as Kekulé vortex\cite{Hou2007,Seradjeh2008}. However, the physical mechanisms inducing such topological defects remain elusive, preventing experimental realisations.
Here, we report the observation of Kekulé vortices in the local density of states of graphene under time-reversal symmetry. The vortices result from intervalley scattering on chemisorbed hydrogen adatoms and have a purely electronic origin. Their $2\pi$ winding is reminiscent of the Berry phase $\pi$ of the massless Dirac electrons. Remarkably, we observe that point scatterers with different symmetries such as divacancies can also induce a Kekulé bond order without vortex. Therefore, our local-probe study further confirms point defects as versatile building blocks for the control of graphene's electronic structure by Kekulé order\cite{Cheianov2009,CheianovPRB2009,Gutierrez2016,Bao2021,Qu2022}.

\section*{Introduction}
\begin{figure*}[t]
	\includegraphics[width=\textwidth]{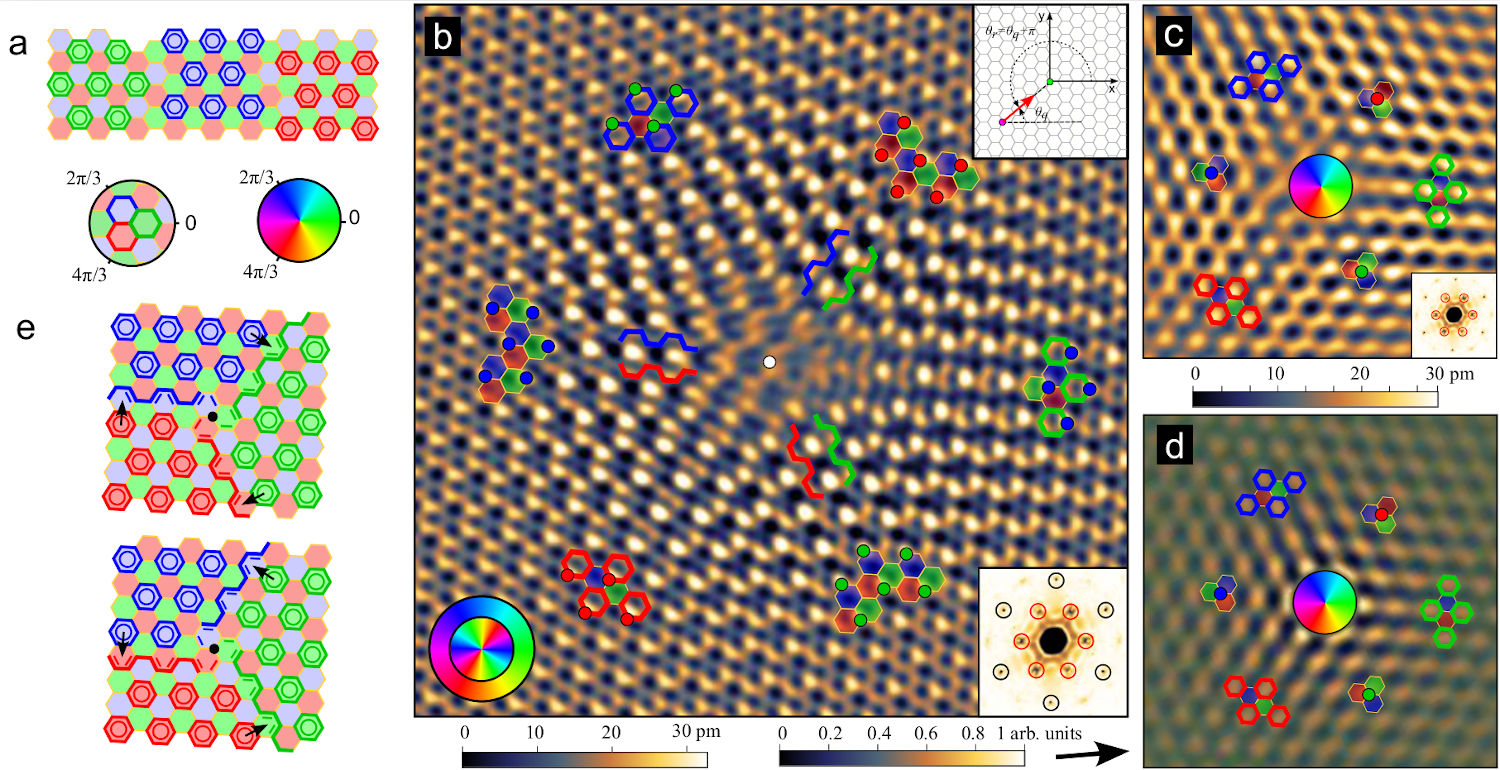}
	 \caption{\textbf{Observation of the Berry-Kekulé vortex induced by hydrogen adatom on graphene}. \textbf{a,} Schematic illustration showing three distinct Kekul\'e orders in graphene. \textbf{b,} STM image of graphene with a hydrogen adatom in the center for a tip bias $V_b$=400 mV and tunneling current $i_t=45.5$ pA. The imaged area of 7.4$\times$7.4 nm$^2$ has been filtered (see Extended Data Fig.~\ref{fig: E Kekulé paving} for raw images) to remove the low frequency signal associated to the H atom which position is given by the white disk. The bottom right inset shows selected harmonics. The colored tiling evidences three domains, each corresponding to one of the three Kekulé orders (bold hexagons) and separated from one another by domain walls along the armchair direction. The outer colour wheel (shown in the bottom left) labels the winding of the Kekulé order parameter around the hydrogen adatom. The coloured dots highlight the usual on-site quasi-particle interference signal\cite{Dutreix2019}, whose $4\pi$ winding is highligted by the central colour wheel (bottom left). The pseudo-spin of incoming electrons scattering off the H atom is locked on the azimutal coordinate $\theta_{\bf r}$ of the STM tip\cite{Dutreix2019} (upper right inset). \textbf{c,} Filtered STM image with the Kekulé harmonics only (see inset) for comparison with the LDOS calculated using the Green's function approach. The imaged area is 4.8$\times$4.8 nm$^2$. \textbf{d,} LDOS around a hydrogen adatom calculated using the Green's function approach and integrated between 0 and 400 meV in order to provide comparison to the experiment. The simulated image includes on-site and bond contributions (see the main text, Extended Data Fig.~\ref{fig: E LDOS th} and Supplementary Information for details).
 \textbf{e,} Clar's sextet configurations of graphene in presence of a hydrogen adatom (black dot) illustrating the emergence of a Kelulé vortex.	\label{fig:Hatom}}
\end{figure*}

Real-space topological defects in crystals exhibit exotic electronic properties\cite{Mermin1979,Nelson2002}, especially when combined with the reciprocal-space topological phase hosted by the bulk\cite{Juricic2012,teo2017topological}. In two-dimensional hexagonal lattices, a vortex in the Kekulé order parameter is of particular interest for charge fractionalisation without breaking time-reversal symmetry\cite{Hou2007,Seradjeh2008}. It is a two-dimensional extension of the charge fractionalisation at domain-wall solitons in the one-dimensional spinless model of polyacetylene\cite{Su1979,Su1980}. 
Introducing the twofold spin degeneracy, a charged topological defect carries zero spin, while the neutral defect carries spin \sfrac{1}{2}. The fractionalisation mechanism is therefore manifested by this unusual spin-charge relation. 
Similar to the dimerisation in one-dimensional polyacetylene, the Kekulé order in graphene corresponds to the $\sqrt{3}\times\sqrt{3}R30^\circ$ unit cell
tripling, with a distinct bond order within one of the three equivalent hexagonal rings. These three degenerate states define an angular order parameter space as shown in Figure~\ref{fig:Hatom}a.
A Kekulé vortex of winding $2\pi$ corresponds to the  alternation of the three Kekulé domains upon encircling a singularity, which could be implemented in optical\cite{Menssen2020,Gao2020} and acoustical\cite{Gao2019,Ma2020} metamaterials. In graphene, recent experiments have demonstrated that a Kekulé bond order emerges from the intervalley coherent quantum Hall ferromagnet states in the zeroth Landau level\cite{Li2019,Coissard2022,Liu2022}. These states can host skyrmionic topological excitations appearing as Kekulé vortices\cite{Goerbig2017,Goerbig2021,Liu2022}. However, these schemes require a strong magnetic field and the Kekulé vortex without breaking time-reversal symmetry remains out of reach. At zero magnetic field, theory shows that a missing electronic site provides a fractionalisation mechanism analogous to that of the Kekulé vortex.\cite{Ovdat2020,Ducastelle2013} Such point defects can scatter electrons from one valley to another, thereby connecting the two valleys and offering a way to stabilize the Kekul\'e bond order \cite{Cheianov2009,CheianovPRB2009,Gutierrez2016,Bao2021,Qu2022}. We now demonstrate that they also induce Kekulé vortices in graphene. 

\section*{Observation of a Kekulé vortex near an H atom on graphene.}

Figure~\ref{fig:Hatom}b shows a scanning tunneling microscopy (STM) image of graphene with a single chemisorbed hydrogen adatom (see Methods for experimental details and Extended data Fig.~\ref{fig: E H3 atom} for another example.). In this image, we have filtered out the intense signal of the close-to-zero-energy state induced by the adatom\cite{Pereira2006,Yazyev2007,Wehling2007}, in order to highlight the result of elastic intervalley scattering of the massless relativistic electrons by the adatom\cite{Brihuega2008}. The image reveals a hexagonal superlattice commensurate with that of graphene but with a unit cell three times larger, as emphasised by the three-colour tiling (See Extended data Fig. 2 for details). The pattern is dominated by the onsite signal on sublattice B (assuming the H adatom is on sublattice A), which alternates between the three nonequivalent B sites. The coloured dots and the inner colour wheel in Fig.~\ref{fig:Hatom}b highlights that this onsite signal winds $4\pi$ when circling around the adatom, consistent with previous studies\cite{Dutreix2016,Dutreix2019,Dutreix2021}. A closer inspection of the interference pattern further reveals the underlying signal on graphene bonds, which is consistent with the Kekulé ordering. In Figure~\ref{fig:Hatom}b, bold coloured hexagons identify three distinct Kekulé domains separated from one another by domain walls along the armchair direction. Remarkably, the Kekulé bond order shows a $2\pi$ winding when circling around the adatom. This demonstrates that the chemisorbed H adatom induces a Kekulé vortex on the surrounding graphene bonds.

\begin{figure*}[t]
	\centering
	\includegraphics[width=\linewidth]{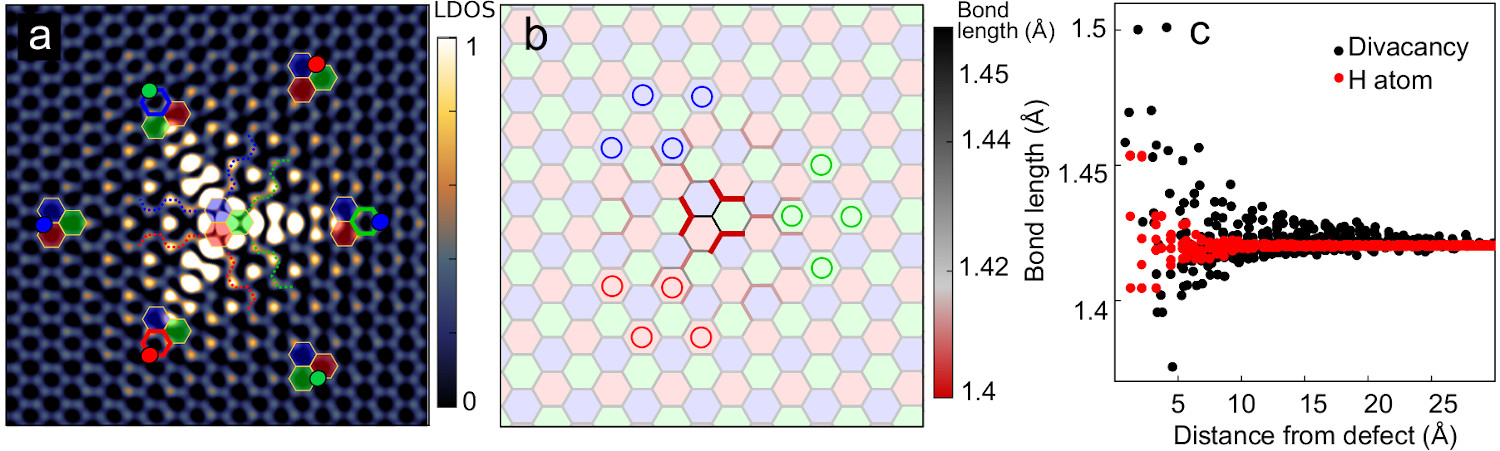} 
	\caption{\label{fig:Kekulé distortion} \textbf{Kekulé vortex induced by the hydrogen adatom from first principles. a,} STM image simulated using DFT with a tiling defined in the same way as in Fig.~\ref{fig:Hatom}b. The calculated LDOS was integrated between 0 and 400 meV as in the experiment. The image area is 4.2$\times$4.2 nm$^2$. \textbf{b,} Relaxed atomic structure of graphene with a hydrogen adatom. The bond lengths are coded as the color and thickness of the bonds. The colored tiling is superposed on the graphene lattice to show the winding of the bond length. The area shown is 3$\times$3~nm$^2$. \textbf{c,} Calculated bond length distribution as a function of distance to the hydrogen adatom and divacancy defect.}
\end{figure*}

\section*{Establishing the electronic origin of the Kekulé vortex}

We will now show that the Kekulé vortex has a purely electronic origin related to the band topology of graphene rather than to defect-induced atomic displacements. The bond ordering of the Kekulé type relates to the electron density between carbon atoms, which are no longer equivalent due to the presence of the hydrogen adatom. Intervalley scattering resolved at the tip position $\vec r$ yields the following bond contribution to the local density of states (LDOS):
\begin{equation}
    \Delta \rho_{AB}(\vec r)\propto \sum_{n=1,2,3} \cos(\Delta \vec K_n \cdot \vec r-\theta_r+\phi_n),
    \label{Eq: Delta rho AB}
\end{equation}
where $\phi_{n}=(2n+1)\pi/3$ and $\Delta \vec K_n$ is a scattering wavevector between the valleys responsible for the hexagonal superlattice pattern (see Supplementary Information). This bond contribution can be understood as a two-path loop interference allowed by the overlap of neighbouring $p_z$ orbitals of carbon atoms.
The vortex originates from the polar angle $\theta_{\vec r}$ indexing the tip orientation, which is known to be a real-space representation of the electron pseudospin $\theta_\vec q$ defined from a Dirac point in momentum space (see inset in Fig.~\ref{fig:Hatom}b and Ref.~\onlinecite{Dutreix2019}). Thus, the bond contribution to the LDOS exhibits a $2\pi$ vortex centered on the adatom as a reminiscence of the topological pseudospin vortex, or equivalently the topological Berry phase, which characterizes the band topology of the massless Dirac electrons\cite{Katsnelson2020}. The usual onsite LDOS modulations $\Delta\rho_{A}(\vec r)$ and $\Delta\rho_{B}(\vec r)$ due to interference also contribute to the STM signal\cite{Dutreix2019}. Adding all contributions leads to the energy-integrated LDOS modulations in Fig.~\ref{fig:Hatom}d, which shows very good agreement with the filtered experimental image shown in Fig.~\ref{fig:Hatom}c. We would like to point out that the experimentally observed vortex is additionally reproduced by both our tight-binding and density functional theory (DFT) calculations (see Methods, Fig.~\ref{fig:Kekulé distortion} and extended data Fig. 5). While the Fourier filtering of experimental data does not allow to separate onsite and bond contributions to the LDOS, this can be readily achieved in calculations (see Extended Data Figs.~\ref{fig: E LDOS th}). We note that the vortices are not affected by the energy integration of the LDOS, since the scattering wave-vectors $\Delta \vec K_n$ are energy independent. Energy-resolved STM images shown in Extended Data Fig.~\ref{fig: E LDOS exp} confirm this conclusion.

The Kekulé vortex in the LDOS derives from an intrinsic topological property of the massless Dirac (that is, chiral in pseudospin~\cite{Katsnelson2020}) wavefunctions scattered by the adatom. Therefore, we name it a Berry-Kekulé vortex. At this point, the following two fundamental questions arise: i) Is the Berry-Kekulé vortex necessarily accompanied by a structural Kekulé distortion that would lead to opening a spectral gap? ii) If not, is the Berry-Kekulé vortex accompanied by zero-energy fractional excitations despite the absence of a spectral gap?

To answer the first question, we perform DFT calculations that intrinsically take into account lattice relaxation effects (see Methods). The simulated STM image (Fig.~\ref{fig:Kekulé distortion}a) reproduces accurately the experiment, and in particular the $2\pi$ vortex.
While the presence of the hydrogen adatom does induce minor lattice distortions that are consistent with the winding of the structural Kekulé distortion (Fig.~\ref{fig:Kekulé distortion}b,c), the results are essentially the same as the ones provided by our analytical description in Eq.~\ref{Eq: Delta rho AB} and tight-binding calculations (Extended Data Fig.~\ref{fig: E TB}) that do not include any lattice distortion effects whatsoever. Furthermore, performing DFT calculation without any relaxation does not affect the presence of the vortex (Extended Data Fig.~\ref{fig: E DFT}). This further confirm the electronic origin of the Berry-Kekulé vortex.

\begin{figure*}[t]
	\centering
	\includegraphics[width=\linewidth]{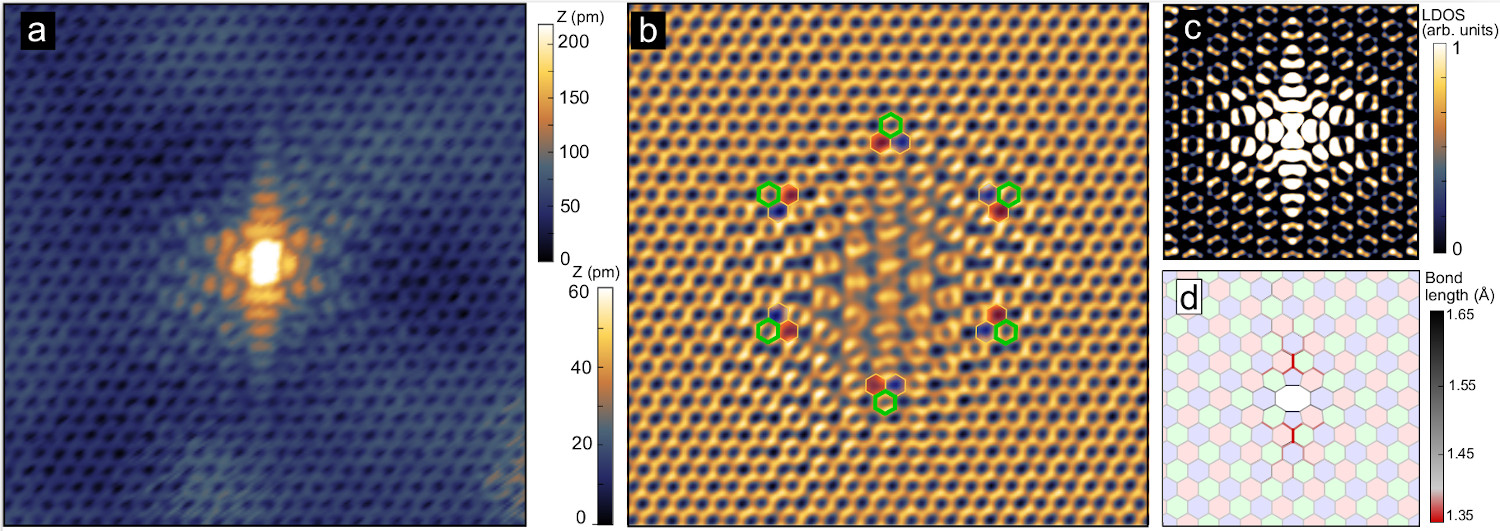} 
	\caption{\textbf{Kekulé order induced by the divacancy defect in graphene. a,} An STM image of a divacancy in graphene ($V_b=500$~mV, $i_t=400$ pA). The image area is 6.2$\times$6.2 nm$^2$. \textbf{b,} Fourier filtered image (same filter as in Fig.~\ref{fig:Hatom}b). \textbf{c,} DFT simulated STM image. \textbf{d,} Atomic structure of the divacancy defect in graphene from DFT calculations. The area shown is 3$\times$3 nm$^2$.}
	\label{fig:Divacancy}
\end{figure*}

Interestingly, the Kekulé vortex can be interpreted in terms of Clar's sextet theory\cite{Clar1983}, a set of empirical rules widely used in the chemistry community that explains graphically the stability of aromatic molecules. The sextet representation depicts the six resonantly delocalized $\pi$ electrons by a circle. Clar's rules state that adjacent hexagons can never be aromatic sextets simultaneously, and that the most stable bond configuration maximizes the number of Clar's sextets. Graphene admits three equivalent resonant Clar configurations corresponding to three Kekulé orders (Fig.~\ref{fig:Hatom}a)\cite{Balaban2009}.
In the presence of a hydrogen adatom that effectively removes one site from the lattice, the circulation of delocalized electrons in the adjacent benzene rings is obstructed, which lifts the degeneracy of Clar's resonant configurations allowed in its surrounding. We then find two possible configurations shown in Fig.~\ref{fig:Hatom}e. The only freedom lies in the positions of the double covalent bonds along the boundaries between two Kekulé domains. Clar's sextet representation leads to the same conclusion: a hydrogen adatom induces a Kekulé vortex in graphene. In supplementary infomation, we formalize the relation between these two pictures.

We will now answer question ii) with regard to fractional excitations accompanying Kekulé vortices, first for spinless electrons. The hydrogen adatom forms a covalent bond with a carbon atom in graphene lattice, eliminating the $p_z$ atomic orbital as a result of rehybridization into the $sp^3$ state of carbon atom\cite{Yazyev2007}. In this sense, a hydrogen adatom is equivalent to a single-atom vacancy in graphene except for stronger structural reconstruction in the latter case. Such a strong potential promotes half a state from the valence band to zero energy, and half a state from the conduction band if the spectrum is particle-hole symmetric. The zero-energy state is fully polarized on the sublattice opposite to that of the removed $p_z$ orbital and presents an algebraic decay due to the gapless relativistic spectrum\cite{Pereira2006,Yazyev2007,Wehling2007}. This quasi-localised state at zero energy has the fractional charge $-e/2$, which corresponds to the charge difference of the occupied valence band with and without the hydrogen adatom\cite{Ovdat2020}. The number of quasi-bound states relates directly to the pseudospin real-space representation $\theta_{\vec r}$ encoded in the scattering wavefunctions\cite{Ovdat2020}. Therefore, the topological winding of the Berry-Kekulé vortex we observe at higher energy and the number of fractional excitations at zero energy are intrinsically intertwined.

Introducing the spin degeneracy doubles the number of zero-energy states, one per spin. At half filling, this implies that one of the two spin-polarised states is fully occupied. Since each bound state is a hybrid superposition of valence-conduction half states of $-e/2$ and $+e/2$ fractional charges, the quasi-bound state around the hydrogen adatom must be neutral with spin $\sfrac{1}{2}$. Remarkably, this very unusual spin-charge relation appears to be supported by two independent experimental observations, one bringing evidence of the neutral charge\cite{Mao2016}, and the other of the magnetic moment\cite{Gonzalez-Herrero:2016aa}. It is also consistent with Lieb's theorem\cite{Lieb1989}. 

\section*{Kekulé order near divacancies}

The existence of the Kekulé vortex and the quasi-bound states induced by the hydrogen adatom defect is related to the symmetries of this scattering center. While a hydrogen adatom breaks the $C_2$ symmetry and sublattice balance of graphene, a divacancy preserves these symmetries. Figure~\ref{fig:Divacancy}a shows an STM image of such divacancy in graphene. The divacancy also induces locally a Kekulé order. Importantly, this Kekulé order is exempt of any winding,
with the Kelulé domain being defined by one of the three orientations of the chemical bond linking the two removed atoms. This is confirmed by DFT calculations (Fig.~\ref{fig:Divacancy}c) which show that the defect also creates significant lattice distortion originating from structural reconstruction due of the formation of two pentagonal rings (Figs.~\ref{fig:Kekulé distortion}c and Fig.~\ref{fig:Divacancy}d). 

A local Kékulé order with or without vortex can therefore be induced in graphene by the specific distribution of atomic defects. Harnessing experimentally these building blocks in a novel type of defect engineering could lead to the long-awaited macroscopic Kekulé order from the cooperative effect of atomic defects\cite{Cheianov2009,CheianovPRB2009}.

\bibliographystyle{nature.bst}
\bibliography{Biblio_Kekule.bib}

\section*{Methods}
\subsection*{Samples and STM measurements}
The samples were grown by thermal decomposition of the carbon-face SiC at temperatures close to 1150$^\circ$ in ultrahigh vacuum.\cite{Varchon2008} Silicon evaporation results in several graphene layers decoupled by rotational disorder. Hydrogen atoms were then deposited by thermal dissociation of hydrogen gas in a custom atomic hydrogen source as described previously.\cite{Gonzalez-Herrero:2016aa} STM images were obtained in the constant current mode in a custom ultrahigh vacuum setup at 5~K. 

\subsection*{Tight-binding calculations}

The nearest-neighbor tight-binding Hamiltonian of graphene is expressed as
     \begin{align}\label{eq-sm-grtb}
        H &= \sum_{\langle i,j\rangle } tc^\dagger_i c_j +h.c. \\ \nonumber &= t\begin{pmatrix}0 & 1+e^{ik_{x}}+e^{ik_{y}}\\
        1+e^{-ik_{x}}+e^{-ik_{y}} & 0
        \end{pmatrix},
     \end{align}
    where the nearest-neighbor hopping integral $t=-2.7$~eV.
The TB calculation is carried out with periodic boundary conditions, with the hydrogen adatom modelled as a large on-site potential ($V=100|t|\approx270\text{eV}$). 
The supercell size of $27\times 27$ unit cells of graphene was used to reduce the spurious effects due to the mutual interference between the periodic images of adatoms.

The information about the phase difference between wavefunction amplitudes on the nearest-neighbor sites, that is the bond order, is needed in order to describe the Kekulé order.
We define the bond parameter as the LDOS between two neighboring atoms $i,j$.
On the basis of TB model, we consider the wave function as the product of the TB eigenvector $\psi$ and an envelope function $f(r)$
    $\phi(r) = \sum_{n} \sum_{i} \phi_i^n f(r-r_i)$,
taking the middle point between the atoms $r_{ij} = (r_i + r_j )/2$ LDOS writes
\begin{align}
    \rho(r_{ij}) &= |f(a_0/2)(\psi_i+\psi_j)|^2 \nonumber \\
    &= f(a_0/2) (2Re\langle\psi_i|\psi_j\rangle + \psi_i^*\psi_i+\psi_j^*\psi_j),
\end{align}
in which the inner product $\langle \psi_i|\psi_j\rangle$ governs the modulation of bond order.
Therefore, we use the orbital overlap $Re\langle\psi_i|\psi_j\rangle$ as the bond-order operators in the TB calculation.
In the presence of a scattering center, $\langle \psi_i|\psi_j \rangle $ is perturbed by the inter-sublattice Green's functions $G_{AB}$ and $G_{BA}$ (See SM\ \cite{SM}), since the atoms connected by the bond are in different sublattices.
The bond order is plotted by integrating the LDOS from 0 to 300 meV. 

\subsection*{DFT calculations}
First-principles calculations were performed using the  SIESTA code~\cite{soler2002siesta}.
We use the double-$\zeta$ plus polarization localized orbital basis set combined with the local density approximation exchange-correlation functional~\cite{LDA_PRB}.
The energy shift for constructing the localized orbital basis functions was set to 275~meV, and the real-space cutoff to 250~Ry.
The structural relaxation was performed using the conjugate gradient  method.

The simulated STM images were produced using the  \textit{plstm} module of the SIESTA package, as a post-processing step following the DFT calculations.
The images were simulated from LDOS at a constant height of 2.5~Bohrs above the graphene plane.

\section*{Acknowledgements}
V.T.R. acknowledges the support from the ANR Flatmoi project (ANR-21-CE30-0029). Y.G. and O.V.Y. acknowledge support from the Swiss National Science Foundation (grant No. 204254). Computations were performed at the Swiss National Supercomputing Centre (CSCS) under projects No. s1146 and the facilities of the Scientific IT and Application Support Center of EPFL.
C.D. acknowledges the support from Quantum Matter Bordeaux and the SMR department under the projects TED and CDS-QM. I.B.  acknowledges the support from the “(MAD2D-CM)-UAM” project funded by Comunidad de Madrid, by the Recovery, Transformation and Resilience Plan, and by NextGenerationEU from the European Union, the Spanish Ministry of Science and Innovation (Grant PID2020-115171GB-I00) and the Comunidad de Madrid NMAT2D-CM program under grant S2018/NMT-4511

\section*{Author contributions}
H.G.-H., M.M.U. and I.B. performed the experiments under the supervision of I.B..  V.T.R. discovered the Kekulé vortex. Y.G. performed DFT and TB calculations under the supervision of O.V.Y. 
Y.G. and C.D. performed Green's function calculations. M.I.K. gave technical support and conceptual advice. Y.G., C.D., O.V.Y. and V.T.R. wrote the manuscript with the input of all authors. V.T.R. coordinated the collaboration.

\renewcommand{\figurename}{\textbf{Supplementary Figure}}
\setcounter{figure}{0}

\begin{figure*}
	\centering
	\includegraphics[width=\textwidth]{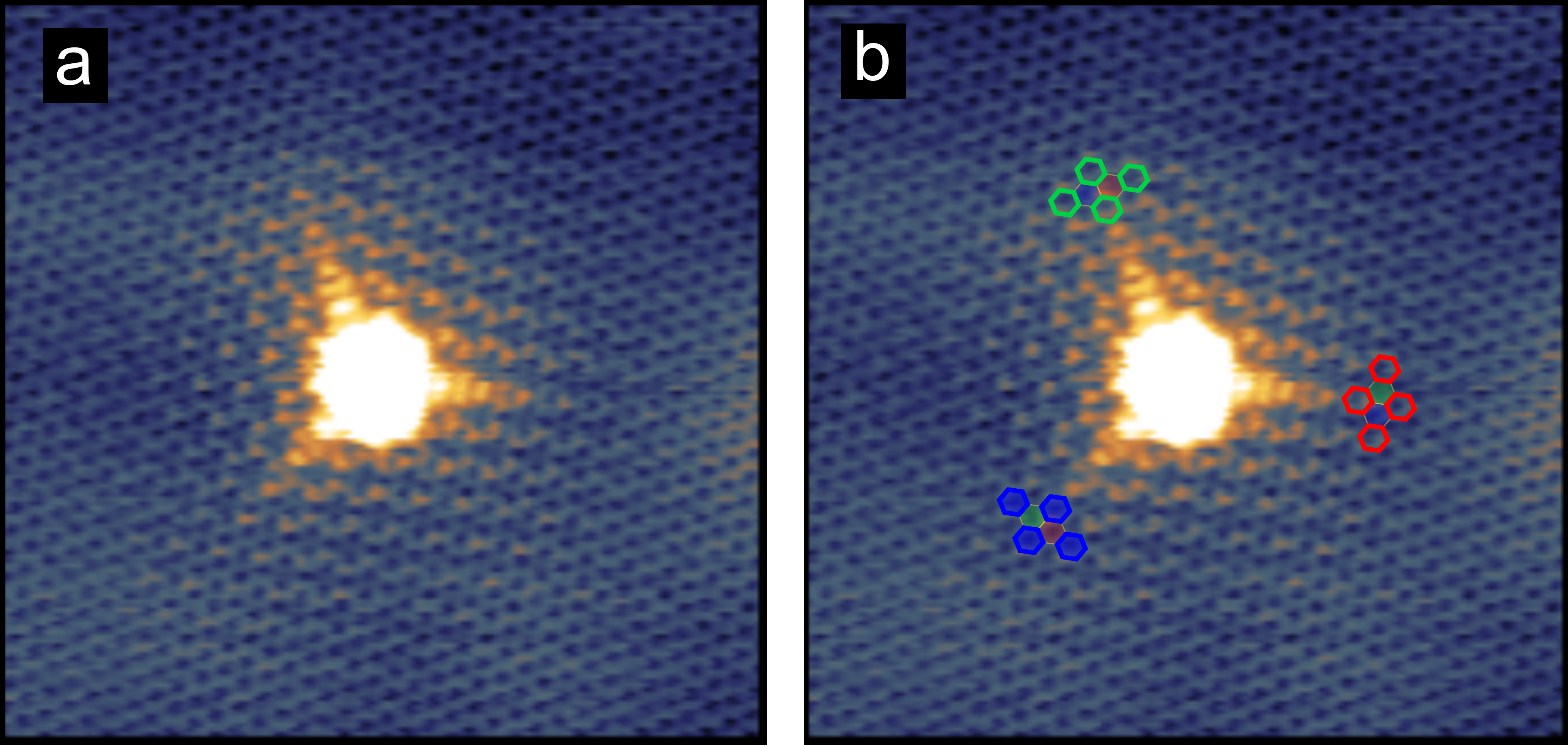}
	\caption[]{\textbf{Kekulé vortex around another H atom.} \textbf{a,} STM image of another H atom measured in an different experimental run. The tunneling conditions and image size are the same as in Fig.~1 of the main text \textbf{b,} Same image with the rgb tiling.}
	\label{fig: E H3 atom}
\end{figure*}

\begin{figure*}[t!]
	\includegraphics[width=\textwidth]{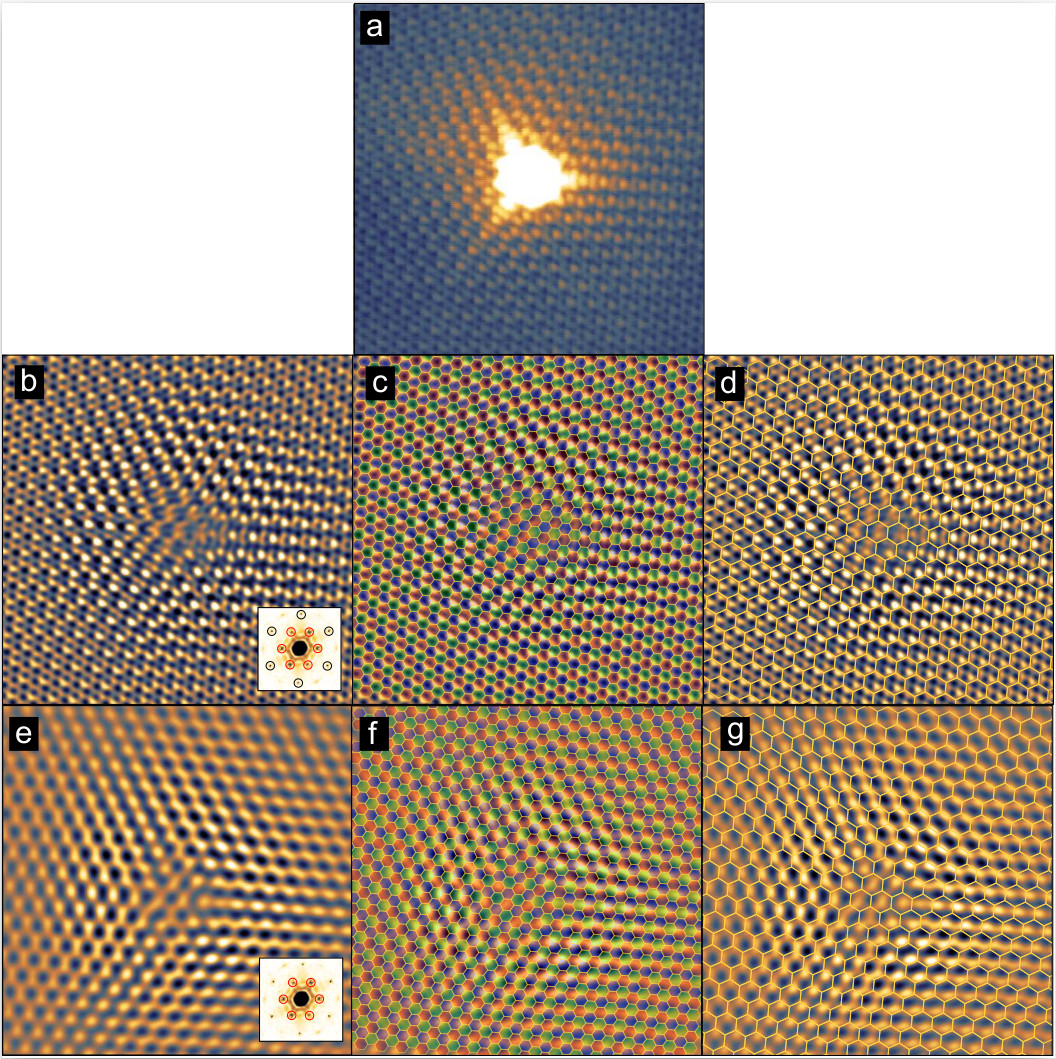}
   	\caption{\textbf{Kekulé paving of the STM image}. \textbf{a,}  Raw STM image of the H atom on graphene from which Fig.~\ref{fig:Hatom}c of the main text is obtained. The Kekulé bond order is seen far from the bright protrusion (the H atom) in the center. The image is 7.4$\times$7.4 nm$^2$ in size and was measured with $V_b$=400 mV and $i_t=45.5$ pA. \textbf{b,} Fourier filtered image including the intervalley scattering and graphene signal (see harmonics selected in the inset). \textbf{c,} Filtered image with a red, blue, green paving. \textbf{d,} Filtered image with a kekulé lattice superimposed.  \textbf{e,} Fourier filtered image including the intervalley scattering only (see harmonics selected in the inset). \textbf{f,} Filtered image with a red, blue, green paving. \textbf{g,} Filtered image with a Kekulé lattice superimposed. }
	\label{fig: E Kekulé paving}
\end{figure*}

\clearpage

\begin{figure*}
    \centering
    \includegraphics[width=\textwidth]{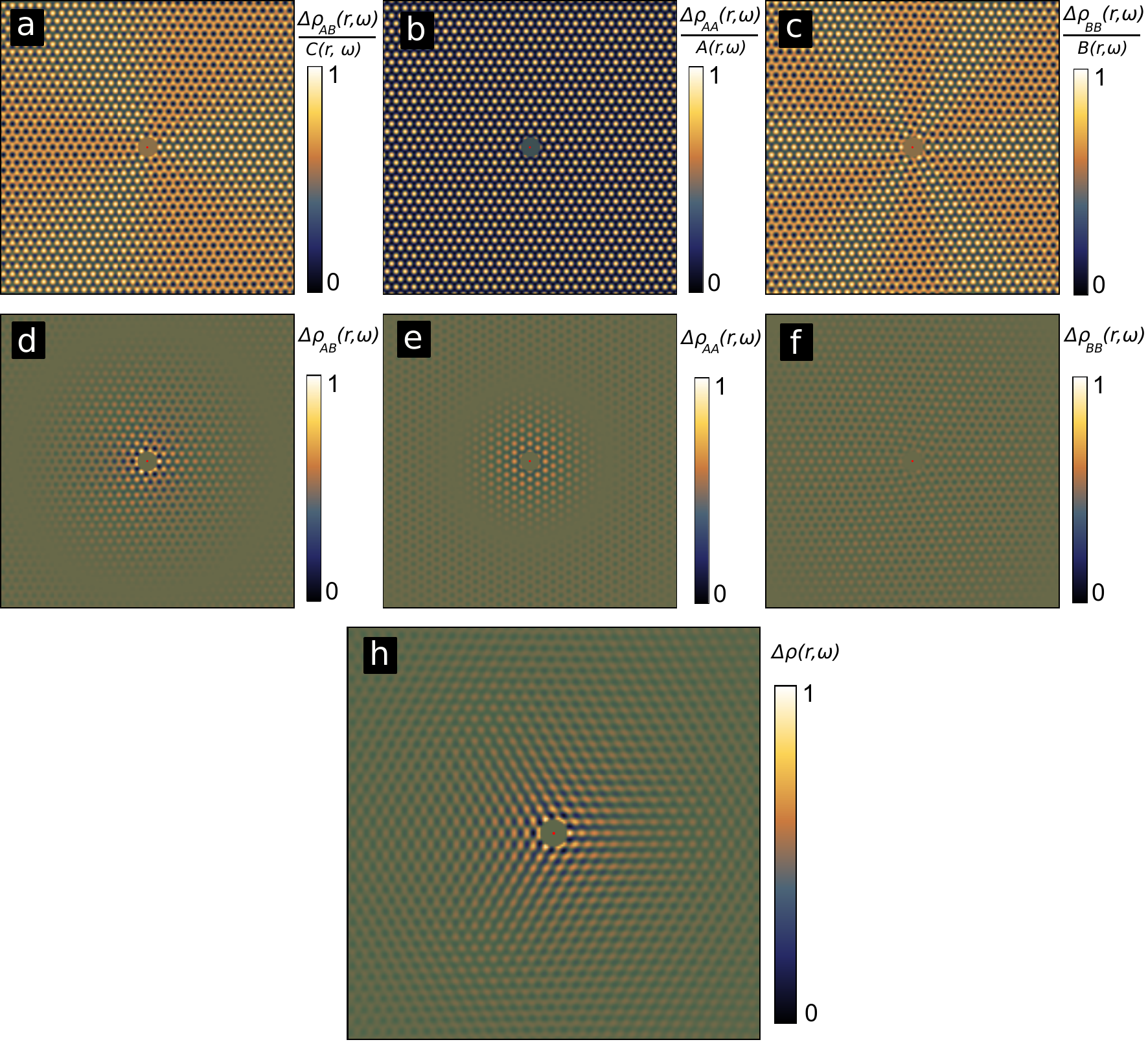}
    \caption[]{\textbf{Different contributions to the LDOS calculated at 150 meV.} \textbf{a,} Image of normalized $\Delta\rho_{AB,BA}(r,\omega)$. The signal contains a $2\pi$ vortex. \textbf{b,} Image of normalized $\Delta\rho_{AA}(r,\omega)$. The signal does not contains a vortex. \textbf{c,} Image of normalized $\Delta\rho_{AB}(r,\omega)$. The signal contains a $2\pi$ vortex. \textbf{d,} Image of $\Delta\rho_{AB,BA}(r,\omega)$. \textbf{e,} Image of normalized $\Delta\rho_{AA}(r,\omega)$. \textbf{f,} Image of normalized $\Delta\rho_{AB}(r,\omega)$. \textbf{h} Sum of the contributions shown in panels \textbf{d, e} and \textbf{f}. The signal is integrated between 0 and $qV_{bias}$ to reproduce the conditions of STM images in the main text. The colormap in \textbf{d, e} and \textbf{f} corresponds to that of \textbf{h,} to highlight the weight of each contribution in the total signal. See the Supplementary Information document for the derivation of each contribution. The image areas are 14$\times$14 nm$^2$.}
    \label{fig: E LDOS th}
\end{figure*}

\begin{figure*}
    \centering
    \includegraphics[width=\textwidth]{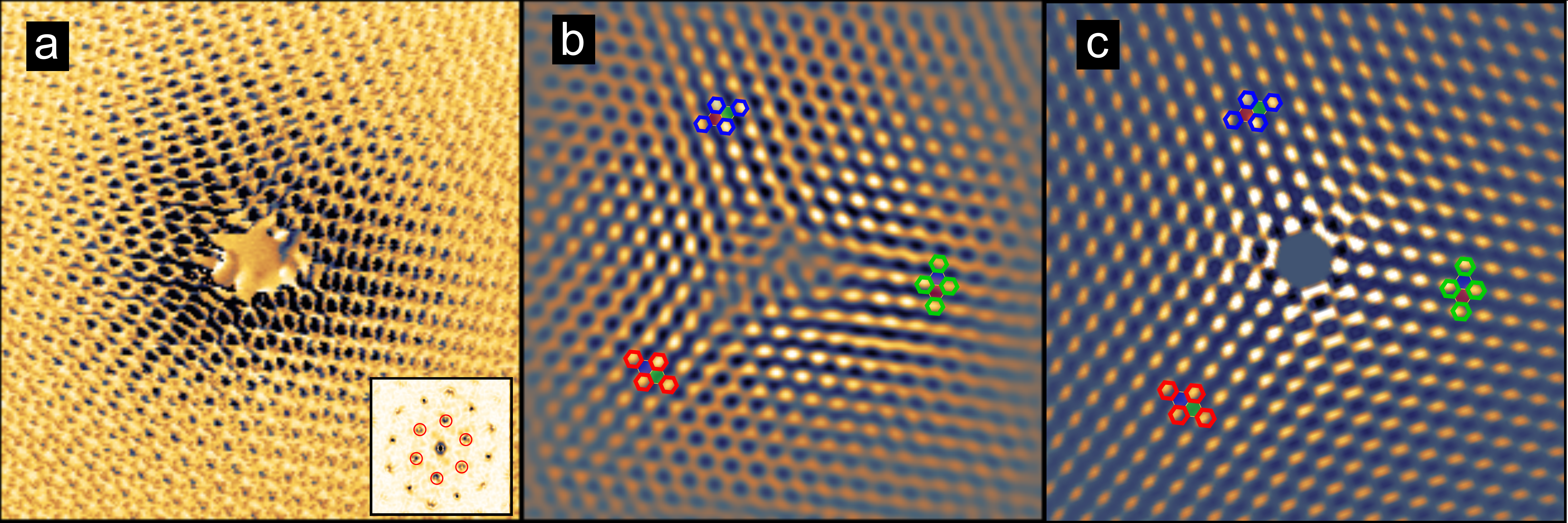}
    \caption[]{\textbf{Energy resolved images of a hydrogene adatom on graphene.} \textbf{a,} Energy resolved STM image measured by phase sensitive detection with a 2~mV rms ac voltage at 830~Hz. The tunneling parameters are $i_t=80$~pA and $V_b=50$ ~mV. The inset shows the Fourier filter used to produce panel \textbf{b}. \textbf{b,} Fourier filtered image allowing to discriminate the three Kekulé orders. \textbf{c,} Total theoretical LDOS calculated at 50~meV as described in the Supplementary Information. }
    \label{fig: E LDOS exp}
\end{figure*}

\begin{figure*}
    \centering
    \includegraphics[width=8.6cm]{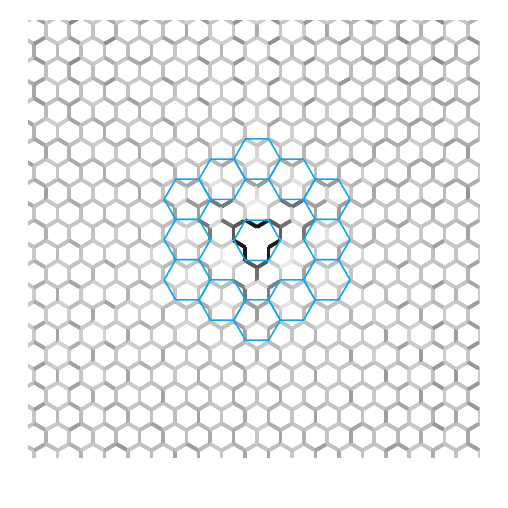}
    \caption[]{\textbf{Bond order at the hydrogen adatom from tight-binding calculations.} The hexagons covering $\sqrt{3} \times \sqrt{3}$ Wigner-Seitz cells serve as a guide-of-eyes to track the evolution of the Kekul\'e order. }
    \label{fig: E TB}
\end{figure*}

\begin{figure*}
    \centering
    \includegraphics[width=\textwidth]{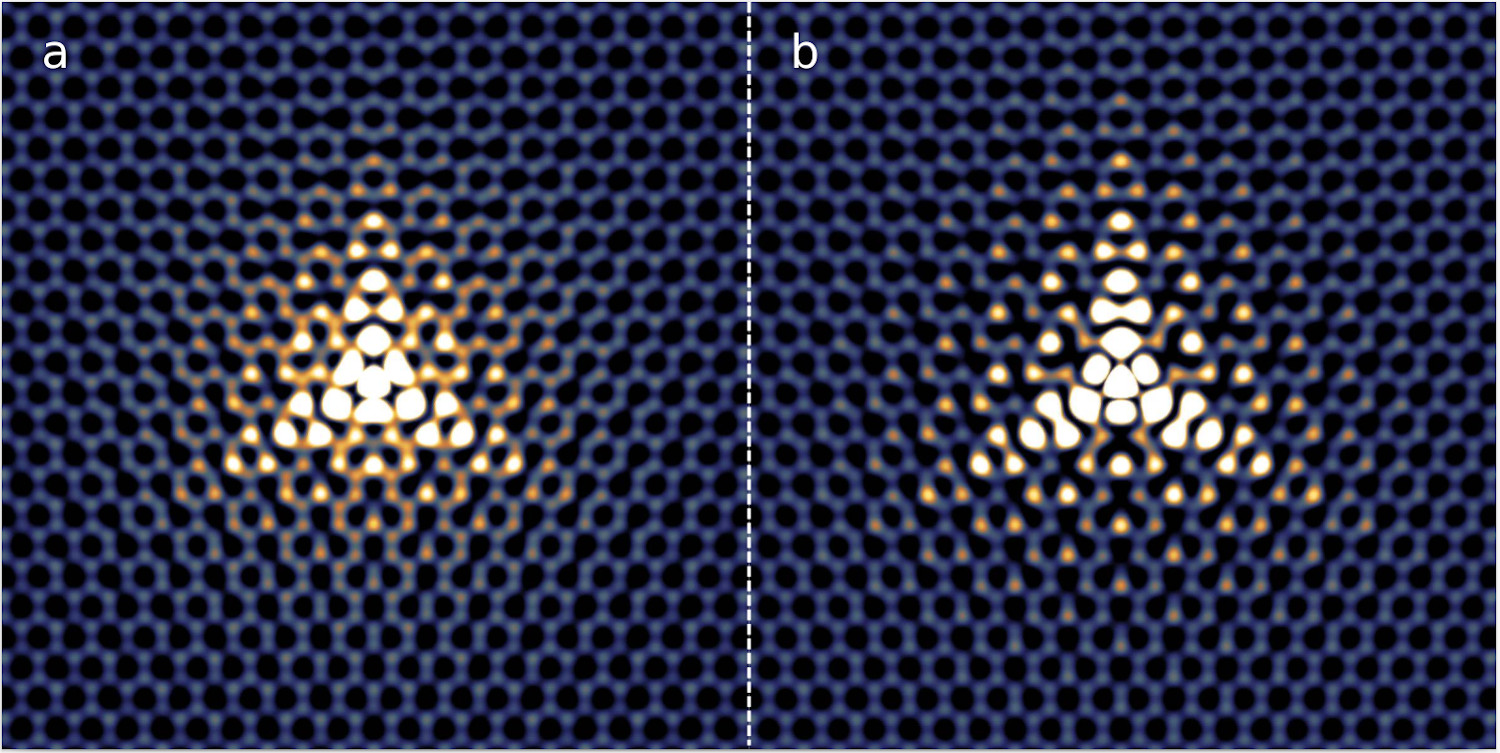}
    \caption[]{\textbf{The effect of structural relaxation on bond order from DFT calculation.} \textbf{a,} STM image simulated by DFT without structural relaxation. \textbf{b,} Same image with structural relaxation.}
    \label{fig: E DFT}
\end{figure*}

\end{document}